\documentclass{tcibook}
\usepackage{fancyhea}
\usepackage{sidecap}
\usepackage{wrapfig}

\usepackage{work}
\usepackage{bm}       
\usepackage{graphicx}
\usepackage{hyperref}      
\usepackage{chngcntr}
\usepackage{hyperref}
\counterwithout{figure}{chapter}

\newcommand{\nc}{\newcommand}  

\def\beq{\begin{equation}}
\def\eeq#1{\label{#1}\end{equation}}
\def\eeqn{\end{equation}}

\newenvironment{Eqnarray}%
   {\arraycolsep 0.14em\begin{eqnarray}}{\end{eqnarray}}
\def\beqa{\begin{Eqnarray}}
\def\eeqa#1{\label{#1}\end{Eqnarray}}
\def\eeqan{\end{Eqnarray}}

\nc{\ra}{\rightarrow}  
\nc{\slsh}{\slash\hspace*{-0.22cm}}
\def\Re{{\cal R \mskip-4mu \lower.1ex \hbox{\it e}\,}}
\def\Im{{\cal I \mskip-5mu \lower.1ex \hbox{\it m}\,}}

\nc{\vev}[1]{ \left\langle {#1} \right\rangle }
\nc{\bra}[1]{ \langle {#1} | }
\nc{\ket}[1]{ | {#1} \rangle }
\nc{\fb}{\,{\rm fb}^{-1}}
\nc{\ev}{{\rm eV}}
\nc{\kev}{{\rm keV}}
\nc{\Mev}{{\rm MeV}}
\nc{\gev}{{\rm GeV}}
\nc{\tev}{{\rm TeV}}
\nc{\mev}{{\rm MeV}}

\def\del{\partial}
\def\Dslash{\not{\hbox{\kern-4pt $D$}}}
\def\dslash{\not{\hbox{\kern-2pt $\del$}}}
\def\pslash{\not{\hbox{\kern-2pt $p$}}}
\def\ETmiss{ \not{\hbox{\kern-4pt $E$}}_T }

\def\msb{{\bar{\ssstyle M \kern -1pt S}}}

\begin{document}

\def\bibname{References}

\bibliographystyle{utphys}  

\raggedbottom

\pagenumbering{roman}

\parindent=0pt
\parskip=8pt
\setlength{\evensidemargin}{0pt}
\setlength{\oddsidemargin}{0pt}
\setlength{\marginparsep}{0.0in}
\setlength{\marginparwidth}{0.0in}
\marginparpush=0pt

\pagenumbering{arabic}
\renewcommand{\arraystretch}{1.25}
\addtolength{\arraycolsep}{-3pt}


\chapter*{Cosmic Visions Dark Energy: Technology}

\begin{center}\begin{boldmath}

Scott Dodelson, Katrin Heitmann, Chris Hirata, Klaus Honscheid, Aaron Roodman, Uro\v{s} Seljak, An\v{z}e Slosar, Mark Trodden


\end{boldmath}\end{center}

\section*{Executive Summary}

A strong instrumentation and detector R\&D program has enabled the
current generation of cosmic frontier surveys. A small investment in
R\&D will continue to pay dividends and enable new probes to investigate the
accelerated expansion of the universe.  Instrumentation and detector
R\&D provide critical training opportunities for future generations of
experimentalists, skills that are important across the entire DOE HEP
program.

\counterwithout{table}{section}
\newpage
\renewcommand*\thesection{\arabic{section}}
\section{Instrumentation and Detector R\&D}

The recent rapid progress in cosmology can be attributed to new experiments inspired by detector technologies developed over the last decade. While ongoing large scale photometric surveys such as the Dark Energy Survey (DES), KIlo-Degree Survey (KIDS), and Hyper SuprimeCam (HSC) produce data on millions of galaxies providing new insights on the accelerated expansion of the universe and the growth of structures, the next generation of experiments, including DESI and LSST, is already under construction.  All of these experiments depend critically on advanced photon detectors such as thick, high resistivity CCDs that provide very high quantum efficiency in the near-infrared, developed at the DOE's Lawrence Berkeley National Laboratory.  Table \ref{tab:CCD} shows some characteristics of current and future wide field cameras enabled by these technological advances. Development of these detectors together with R\&D on robotically actuated fiber positioners also enable a new class of multi-object spectroscopic surveys including DESI, that will go on the sky later this decade and provide significant improvements in throughput, number of objects and resolution (Table \ref{tab:MOS}). In this white paper we review the status of ongoing or planned  instrumentation and detector R\&D efforts and discuss how these can enable the next generation of dark energy experiments.
We start by defining the goals for a successful Cosmic Vision Dark Energy R\&D program:\
\begin{itemize}
\item Enhance detector characteristics or develop new devices that allow the extraction of more information from cosmological surveys.
  \begin{itemize}
    \item Improve quantum efficiency over a wider wavelength range.
    \item Eliminate or reduce readout noise and develop technologies to suppress other backgrounds.
    \item Increase survey efficiency, e.g. by reducing readout time.
  \end{itemize}
  \item Develop devices and instrumentation to enable new dark energy probes such as 21 cm experiments or efficient measurements of the velocity field.
  \item Develop the technology needed to calibrate the instruments to reduce systematic uncertainties.
  \item Provide training for the next generation of experimentalists.
\end{itemize} 

Much research is also needed on the development of advanced data processing and reconstruction algorithms. This important topic, however, is outside the scope of this report.

\begin{table}[h]
\small
\begin{center}
\begin{tabular}{lll}
\hline & Dark Energy Camera & LSST Camera \\ \hline 
Thickness [$\mu m$] & 250 & 100 \\ \hline
Pixel Size [$\mu m$] & 15 & 10 \\ \hline
Readout Noise [$e^- RMS$] & $<15$   & $<9$ \\ \hline 
Readout Time [sec] & 20 & 2 \\ \hline
Amplifiers/Sensor & 2 & 16 \\ \hline
CCD pixel format & 2k x 4k & 4k x 4k \\ \hline
Number of Science CCDs & 62 & 189 \\ \hline   
Number of Filters & 5 & 6 \\ \hline
Field of View [sq. deg.] & 2.2 &  9.6 \\ \hline 
\end{tabular}
\caption{Current and Future Wide Field Cameras}
\label{tab:CCD}
\end{center}
\normalsize
\end{table}

\begin{table}[h]
\small
\begin{center}
\begin{tabular}{lllll}
\hline & BOSS/eBOSS & DESI & PFS & 4Most \\ \hline 
Number of Fibers & 1000 & 5000 & 2400 & 2400 \\ \hline
Fiber Size [arcsec] & 2 & 1.4 & 1.1 & 1.4 \\ \hline
Wavelength Coverage [$\mu$m] & 0.36 - 1.0 & 0.36 - 0.98 &0.38 - 1.26 & 0.4 - 0.9 \\ \hline
Resolution & 2000  &2000 -5100 & 2000 - 5000 & 5000 - 18000 \\ \hline
Reconfigure Time [sec] & 900 & 120  & & 480 \\ \hline
Detector Type, Format &  CCD, 4k x 4k & CCD, 4k x 4k & CCD, 2k x 4k & CCD, 6k x 6k  \\
 & & & HgCdTe, 4k x 4k & \\ \hline
Number of Detectors & 4 & 30 & 20 & 8 \\ \hline   
Telescope Mirror [m] & 2.5 & 4 & 8.2 & 4 \\ \hline
Field of View [sq. deg.] & 7.1 &  8.0 & 1.3 & 4.1\\ \hline 
\end{tabular}
\caption{Current and Future Multi-Object Spectrographs}
\label{tab:MOS}
\end{center}
\normalsize
\end{table}


\section{Current Instrumentation and Detector R\&D for Future Dark Energy Experiments}

In this section we present a selection of instrumentation research relevant for future dark energy experiments. The efforts fall into several categories: improved photon detectors, improvements for spectroscopic surveys, and technologies for new experimental probes. This program is well matched to the science questions discussed in the accompanying Science White Paper. 
For details on the major engineering and development efforts supporting the Stage IV dark energy experiments (LSST, DESI) and space missions (Euclid, WFIRST) currently under construction, we refer the reader to the ample technical documentation available on the collaborations' web sites. Details about each of the R\&D efforts follow.




\subsection{CCDs and CMOS Sensors}

Silicon CCDs are the mature technology for ultra-violet to near-infrared wavelengths (300 to 1000 nm).  Their uniform spatial response, excellent linearity, and low read noise at low read rates ($<1$ electron in some current devices) are still unique.  Additionally, they are relatively inexpensive compared to hybrid pixel devices, especially for large sizes of up to 100~cm$^2$.  Specialized formats that may be required for a particular instrument can be designed and fabricated on a quicker time scale than Complementary Metal-Oxide Semiconductor (CMOS) sensors.  The trend to larger telescopes with their great light collection capabilities pushes sensors  to have high frame rates while maintaining low read noise. Historical CCD shortcomings of low frame rates and baroque operating voltages are being addressed with a larger number of readout ports (e.g., 16 for LSST sensors and up to 192 on some LBNL devices) and integrated circuits (e.g., LSST and LBNL).  There is room for further development on both fronts in order to maintain CCDs' good imaging characteristics.  The extreme is represented by work underway to instrument each CCD column with its own readout.  Other ongoing research efforts include reducing readout noise with improved output transistor designs, constructing multi-layer multi-color sensors, developing single photon counting applications through charge multiplying output registers. and extending high quantum efficiency on both the red and blue end of the spectrum. 

CCDs remain the workhorse detector for astronomical applications but for consumer products monolithic CMOS image sensor (CIS) technology has developed into a \$10B/yr market with over 100M device shipments per year, driven largely by mobile phone and tablet camera applications. The driving requirements for the mainstream market include acceptable sensitivity and dynamic range across the visible spectrum, operation over the commercial temperature range 0 to +70C, ``camera-on-a-chip'' integration of sensing and processing, combination of still photo and video capability and low cost per megapixel. Most of these features are not well matched to the needs of scientific experiments and only a small fraction of the industry targets high-performance, scientific imaging. The main application areas are bio-imaging, low-light (surveillance) imaging, and astronomy. As with consumer applications, the primary advantages of CIS over CCDs are low power dissipation and the flexibility afforded by integrating sensors with processing electronics on the same chip. Scientific CMOS imagers have undergone intensive development over the past 10 years and the best devices have performance matching, and in some cases exceeding, that of CCDs. For example, cameras with a 2k$\times$2k CMOS sensor (CIS2020) having 1 $e^-$ read noise at 100 frames per second and 30,000 $e^-$ full well are commercially available. Recent research papers claim read noise as low as 0.5 electrons. Despite this, all current and planned large astronomical instruments (ground- and space-based) are using CCDs for optical and NIR wavelength imaging. CIS have not yet achieved the broadband, high quantum efficiency and dynamic range available in the best CCDs, although there are no intrinsic factors that would prevent them from reaching such performance. In addition, scientific CMOS developers have targeted single-chip instruments, and have therefore not yet addressed the issues involved in fabricating the large chip size, high fill factors, and special packaging required for efficient multi-chip focal planes. 

Hybrid CMOS sensors are a separate, mature technology that has widespread application in infrared astronomy. These devices consist of a photodiode array bump-bonded to a CMOS readout ASIC. Most commonly the photodiode array is fabricated on low-bandgap semiconductors such as HgCdTe which can be sensitive to wavelengths longer than that of silicon. Despite their dominance in infrared focal planes, hybrid CMOS performance at UV and visible wavelengths has been inferior to that of monolithic CIS and CCDs. 
The fast frame rate, shutterless operation, and low noise of CIS make them well suited for fast adaptive optics, guiders, and similar applications. They are already becoming competitive with electron-multiplying CCDs for these applications, which are important for general astronomy. For future cosmic frontier experiments, the exceptionally low noise of scientific CMOS technology may be of most benefit to spectrograph applications, where (unlike in wide field imagers) the need for high dynamic range and high fill factor is less important. However, development of custom CIS devices suitable for a large future imaging DE experiment will require finding a foundry partner with the resources to attach the issues of large chip size, broadband wavelength response, and buttable packaging.

\subsection{CCD Characterization}

Thick deep depletion CCDs, with a large aspect ratio between thickness and pixel size, exhibit a number of effects which complicate the relation between incident photon position and the resulting signal.  Beyond diffusion, these effects include the intensity-dependent PSF effect, or brighter-fatter effect, CCD tree-rings, edge distortions and intrinsic effective pixel size variations.  As a measure of the systematic impact of these CCD phenomena, the brighter-fatter effect must be understood and corrected at the 10\% level for DES galaxy shape measurement and weak lensing, and likely at the 1\% or better level for LSST.  Moreover, the presence of {\it baked-in} spatial systematics can potentially feed through to the final measured power spectra.   A significant amount of R\&D has now been devoted to understanding these effects, as present in DECam and expected in the LSST camera, and have been the subject of two recent workshops held at BNL.  While much has been learned, further characterization is ongoing to fully understand the relevant CCD behavior and how best to account or correct for it.  Relevant work includes the study of on-sky data, laboratory data with artificial sources as well as detailed CCD simulations.  

\subsection{Germanium CCD Detectors}

While silicon is the material of choice for visible imaging, it is insensitive to radiation with wavelength beyond approximately 1 $\mu$m.  For imaging in the short-wave infrared (SWIR) band the dominant technology consists of InGaAs diode arrays bump-bonded to silicon readout integrated circuits and back-illuminated through the InP substrate.  Compared to silicon, these detectors have multiple disadvantages.  They are expensive, fragile, and limited to 150-mm diameter or smaller substrates.  Bump bonding limits the minimum pixel pitch to about 15 $\mu$m.  Furthermore, back-illumination is performed on the chip- rather than wafer-level, and response cuts off below 920 nm due to absorption in the InP substrate.  Finally, the largest InGaAs devices are limited to 1.3 Mpixel and are only approximately 3 cm$^2$.  Larger arrays based on HgCdTe are available and cover a larger portion of the infrared but require substantial cooling and suffer from low yield.

A group at MIT Lincoln Laboratory has recently begun to develop germanium CCDs. Germanium is commercially available in wafer diameters up to 200-mm and can be processed with the same tools used to build silicon imaging devices.  A germanium CCD offers response through the visible and SWIR bands with a cutoff of approximately 1.7 $\mu$m (the same as InGaAs).  The backbone of a CCD is the metal-oxide-semiconductor (MOS) capacitor.  The Lincoln Lab team optimized germanium MOS capacitors to yield the high-quality metal-oxide interface required for low noise CCDs.  Based on these measurements and analysis of other devices, they conservatively project that a germanium CCD will exhibit dark current noise below 100 pA/cm$^2$ at an operating temperature of 150K.  All other sources of noise should be comparable to silicon CCDs. 

The first germanium imagers, with formats up to 1 MPixel with an 8 $\mu$m pixel pitch are currently being designed.  First devices will be fabricated in 2016.  These early chips will be frontside-illuminated and therefore exhibit fairly low sensitivity.  The roadmap for 2017 to 2020 includes development of larger CCDs and the fabrication of back-illuminated devices with high efficiency. Over the longer term, these efforts can lead to germanium CCDs with format, pixel pitch, and capabilities that match current silicon devices.

\subsection{Multi-Color CCDs}

While CCDs are sensitive to a wide range of wavelengths, images are taken in relatively narrow wavelength bands.  Wide-field imagers have typically used $100-150 nm$ wide bands.  Thus most light is actually filtered away in every image.  In some instruments it is possible to multiplex light of different wavelengths, through the use of dichroics and multiple optical paths, but this is not practical for large wide-field imagers.  Alternatively, because the absorption depth in Silicon is highly wavelength dependent, it is possible to construct a sensor that can simultaneously collect multiple wavelengths.  In this scenario, the sensor must be constructed in multiple layers, with the first layer collecting the blue end of the spectrum, with a very short absorption length, and the last layer collecting the red end, with its long absorption length.  The major challenge of this approach is to develop the multiple layers necessary to achieve a multi-color CCD.  One sensor maker, Foveon, developed a three layer CMOS device, used in Sigma SLR cameras. However the noise, quantum efficiency, and band pass in these devices are not adequate for astronomical instrumentation.

\begin{figure}[htbp]
\centering
\includegraphics[width=0.48\columnwidth]{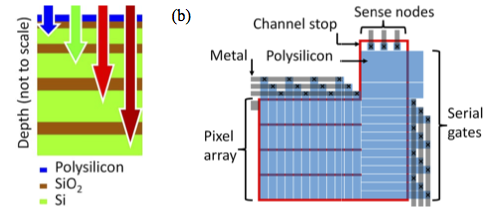}
\includegraphics[width=0.48\columnwidth]{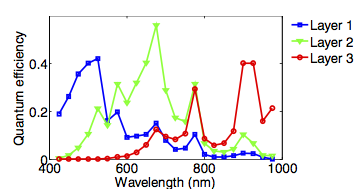}
\caption{Multi-band CCD schematic (left) and simulated quantum efficiency per layer (right)}
\label{fig:multiccd}
\end{figure}

Recently, the SLAC instrumentation group working at the Stanford Nanofabrication Facility developed a proof-of-principle multiband CCD, consisting of semiconductor-on-insulator substrates and multiple device layers.  This sensor builds on the 3D Silicon sensor R\&D conducted for a highly radiation resistant Silicon tracking device.  This initial demonstration device consists of a single pixel three-layer device fabricated by direct wafer bonding and precision grinding.  A schematic of the device and the simulated quantum efficiency per band are shown in Figure~\ref{fig:multiccd}.  The fabricated device successfully demonstrated multi-band light collection and charge extraction.  One interesting feature of this R\&D effort is that the QE per layer is dependent on both the Silicon absorption and substantial interference at the layer boundaries, which could conceivably be exploited to yield more uniform and well separated bands.  Still, these devices will require substantial effort to yield sensors suitable for astronomical images.

\subsection{MKIDs}

Microwave Kinetic Inductance Detectors (MKIDs) are an emerging new detector technology for imaging, photon counting, and low resolution spectroscopy. MKIDs count single photons and measure their energy without read noise or dark current, and with nearly perfect cosmic ray rejection. Their usefulness for astronomical instrumentation has been demonstrated in over 30 observing nights with the ARCONS instrument on the Palomar 200” and the Lick 120” telescopes.

\begin{figure}[htbp]
\centering
\includegraphics[width=0.6\columnwidth]{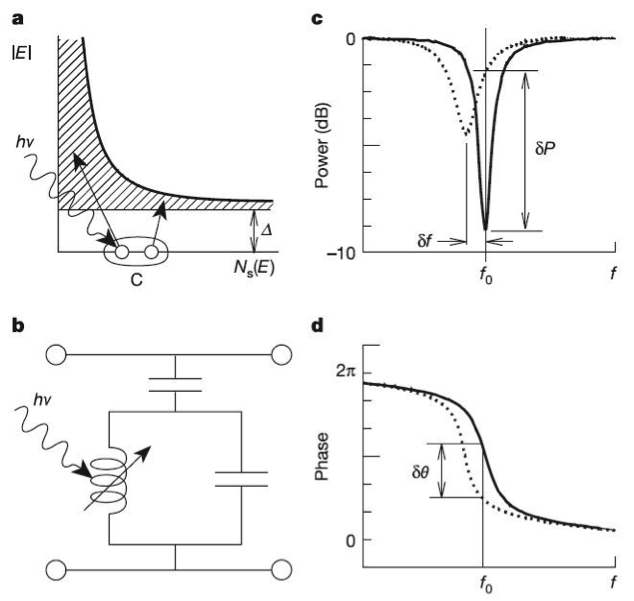}
\caption{The basic operation of an MKID [2]. (a) Photons are absorbed in a superconducting film, producing a number of excitations, called quasiparticles. (b) To sensitively measure these quasiparticles, the film is placed in a high frequency planar resonant circuit. The amplitude (c) and phase (d) of the microwave excitation signal sent through the resonator. The change in the surface impedance of the film following a photon absorption event pushes the resonance to lower frequency and changes its amplitude. If the detector (resonator) is excited with a constant on-resonance microwave signal, the energy of the absorbed photon can be determined by measuring the degree of phase and amplitude shift. }
\label{fig:mkid}
\end{figure}

MKIDs work on the principle that incident photons change the surface impedance of a superconductor through the kinetic inductance effect, see Figure~\ref{fig:mkid}. The kinetic inductance effect occurs because energy can be stored in the supercurrent (the flow of Cooper Pairs) of a superconductor. Reversing the direction of the supercurrent requires extracting the kinetic energy stored in it, which yields an extra inductance term in addition to the familiar geometric inductance. The magnitude of the change in surface impedance depends on the number of Cooper Pairs broken by incident photons, and hence is proportional to the amount of energy deposited in the superconductor. This change can be accurately measured by placing a superconducting inductor in a lithographed resonator.

A microwave probe signal is tuned to the resonant frequency of the resonator, and any photons that are absorbed in the inductor will imprint their signature as changes in phase and amplitude of this probe signal. Since the quality factor Q of the resonators is high and their transmission off resonance is nearly perfect, multiplexing can be accomplished by tuning each pixel to a different resonant frequency with lithography during device fabrication. A comb of probe signals is sent into the device, and room temperature electronics recover the changes in amplitude and phase. MKIDs have the potential of reaching spectral resolution of $R \sim 100$ in the blue edge of the visible spectrum; the current performance achieved in these sensors is $R \sim 15$. Significant R\&D is needed to improve this performance.

A major benefit of an instrument with an MKID focal plane is low resolution spectroscopy of a large number of objects resulting in significantly improved redshift measurements than typically achievable with a multi-filter photometric survey such as DES or LSST.  Improved photometric redshifts can also be obtained using cameras with a large number of narrow band filters, a method currently pioneered by the PAU (Physics of the Accelerating Universe) and JPAS (Javalambre Physics of the Accelerating Universe Astronomical Survey) collaborations. A few people have looked at options to equip the LSST camera with a different set of filters after completion of the baseline survey. Options studied include a larger set of narrow band filters or a set of 6 filters that match the current LSST filters but with shifted wavelength coverage. Such a program would follow the 10 year LSST survey. However, should further studies demonstrate the value of these additional colors for photometric redshift calibration, it could be of interest to obtain the additional data sooner using sets of filters matched to the LSST specifications on other instruments such as the Dark Energy camera on the Blanco telescope.

A workshop on cosmology with low resolution spectroscopy will be held at the University of Chicago in early 2016 and further studies are needed to assess the impact the instruments can have on dark energy science.


\subsection{Instrumentation for 21 cm Experiments}

There are two basic experimental designs for 21 cm cosmology experiments: interferometers and imagers. They share the basic building blocks of antennas and radio feeds and amplifiers which are well understood and commoditized thanks to the telecommunications industry. The requirements on these elements are modest and not pushing the envelope of the technology.

Interferometers rely on fast digital processing to synthesize radio telescope beams in software. This is typically done with custom designed Field-Programmable Gate Array (FPGA) boards or Graphics Processing Units (GPU). Getting the throughput of these signals to match the bandwidth still presents a challenge and will continue to do so, as the number of antennas increase although the advances in computer technology seem likely to keep pace with this trend. Future work in this area can improve the integration between networks and processors to make much more scalable solutions with reduced burden on slow CPUs to handle transactions.  Another major new direction the community is interested in is a new style of correlator which, instead of cross-correlating every pair of antenna elements (for a computational scaling of $O(N^2)$ with number of antennas), grids the electric field incident on an array of antenna elements, and uses a spatial Fourier Transform to allow correlations to scale as O(N log N). Efforts in this area have so far made only modest progress because the complexity of integrating calibration systems into the real-time operation of a correlator.  However, this technology would be a game-changer for radio astronomy and many other applications of radio interferometry.

In imagers, an antenna performs the role of the correlators in interferometers. The immediate advantage is that the expensive digital signal processing technology is not required, but this comes at a very large cost to the immediate field of view. The Green Bank Telescope where the cosmological 21-cm signal was detected for the first time in cross-correlation is an imager.  The compactness of the camera is another advantage of an imaging experiment, allowing receivers to be cooled and achieve better noise properties, leading to 5-25 times faster integrations. Controlling systematics might also be easier with imagers since it is easier to achieve phase stability in hardware and there is no need to deal with noise emitted by amplifiers in individual horns producing spurious correlations. Nevertheless, further work is needed to fully develop the radio imager concept and to assess its value for cosmological measurements.

The field of 21 cm cosmology is well matched to the current expertise in the DOE national laboratory complex. First, accelerator and light-source know-how involves deep understanding of radio-frequency technology which is directly transferrable to this field. Second, the data-rate from these instruments is extremely large and therefore well matched to the high throughput computing expertise that has been developed to support HEP experiments. These set of challenges also offer an excellent opportunity to collaborate with other offices, e.g. ASCR.

\subsection{Ring Resonators}

\begin{figure}[htbp]
\centering
\includegraphics[width=0.95\columnwidth]{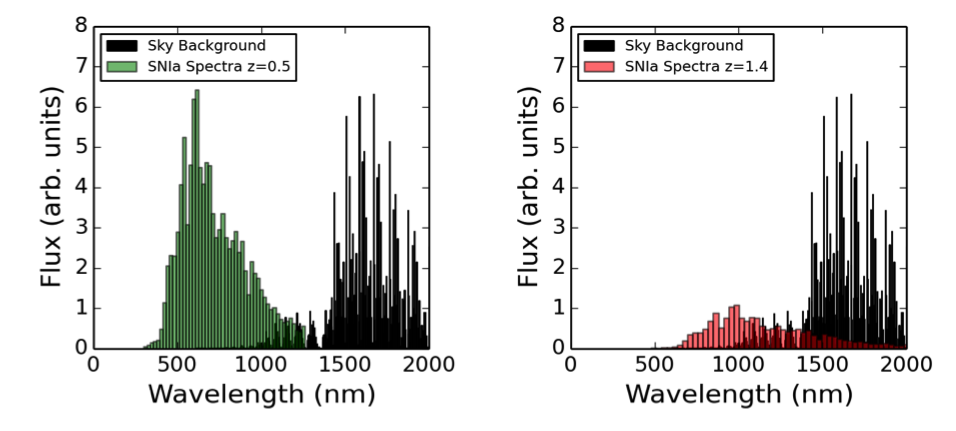}
\caption{Spectra of the optical and NIR wavelengths; a supernova spectrum at redshift $z=0.5$ (left) is compared to the sky background spectra (black spikes), the same comparison is made with a supernova at redshift $z=1.4$ (right)}
\label{fig:ring-wave}
\end{figure}

The current state-of-the-art instrument for SN Type Ia cosmology is the Dark Energy Survey which will record ~3k supernovae.   The LSST will record $\sim$50k supernovae in its deep-drilling fields and with its improved calibration systems will reduce the dominant systematic uncertainty significantly.  However, the next largest expected uncertainty, due to host galaxy dust obscuring the supernova brightness, will become a limiting factor and the LSST instrument is not a significant improvement in addressing this.  Follow-up measurements in the near-infrared are crucial to reduce the dust systematic uncertainty. In addition, recent measurements have shown that a single rest-frame measurement in the near-infrared provides an excellent standard candle.
Ground-based astronomy in the near-infrared region is plagued by the emission lines of the hydroxyl molecule OH in Earth's upper atmosphere.   Figure~\ref{fig:ring-wave} shows spectra of the optical and NIR wavelengths; on the left a supernova spectrum at redshift $z=0.5$ is compared to the sky background spectra (black spikes), on the right the same comparison is made with a supernova at redshift $z=1.4$.

The spikes are actually a collection of $\sim$1000 narrow doublets, with doublet separation of 0.1-0.2 nm, and clean spaces of 5-10 nm between the spikes.  In order to reach the same signal-to-noise level in the NIR as in the optical, one has to integrate 1800 times longer due to the larger (spikes) background.  A group at Argonne National Laboratory has started to investigate the use of ring resonator filters, commonly used in telecommunications, to surgically suppress the narrow lines, while maintaining good efficiency for the remaining wavelengths. These filters would be inserted on the optical fibers that run from the imaging focal plane of a telescope to a NIR spectrograph.

\begin{figure}[htbp]
\centering
\includegraphics[width=0.5\columnwidth]{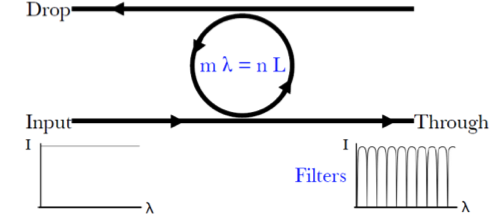}
\caption{Ring Resonator Schematic}
\label{fig:ring-schem}
\end{figure}

Figure~\ref{fig:ring-schem} shows a schematic view of a ring resonator device. The input straight waveguide couples to the ring, and wavelengths that satisfy the resonance conditions, $m\lambda = nL$ (with L being the circumference and n the effective index of refraction of the ring), are coupled to the second straight waveguide and dropped. The net effect is that the resonant wavelengths are suppressed at the output of the initial waveguide. The development of the ring resonator fabrication process has begun, using the facilities at Argonne's Center for Nanoscale Materials. Rings and waveguides have been successfully fabricated; the focus of the process development has now turned to the efficient coupling of the wafer to an optical fiber.  The next steps include a bench test of a 6-ring device followed by on-sky tests performed in collaboration with a team from the Australian Astronomical Observatory. The ultimate goal is the development of a single chip device with hundreds of rings suppressing OH wavelengths, linked in series with a single input and output optical fiber, that will be added to a southern hemisphere instrument and enable inexpensive, near-infrared follow up of LSST supernovae.

Ring resonators are one of a number of technologies under development to suppress OH skylines. A prototype system using optical fiber devices called fiber Bragg gratings has been tested on a limited number of lines, but will be difficult to scale. R\&D on scalable technologies, such as special material-science techniques like metamaterials or silicon-based photonics, is needed to provide an effective general-purpose OH suppression system.

\subsection{Adaptive Optics}

Multi-Conjugate Adaptive Optics systems (MCAO) expand the standard adaptive optics approach by using multiple natural or laser guide stars with several deformable mirrors. MCAO systems can achieve uniform image correction on a field of view significantly larger than the natural isoplanatic patch. The GeMS instrument at Gemini South achieves diffracted limited images over an $85''\times85''$ square field. These fields are too small for cosmological surveys and the required large number of galaxies but these instruments can be of significant value for SN cosmology in particular when combined with OH suppression technology as described in the previous section. By reducing the width of the point-spread function of the joint atmosphere + telescope + detector optical system, adaptive optic systems can yield both substantially improved angular resolution along with significant improvements in the signal-to-noise ratio. An improvement from a seeing of 0.7'' to 0.5'' yields a factor of 2 improvement in the ratio of enclosed energy of the source to the background emission.  

Multi-conjugate adaptive optics systems have been deployed in recent years and further development and refinement of this technology will significantly increase the ability of telescopes to effectively explore the observer-frame near-infrared.  MCAO is critical to SN cosmology, because it allows for fields that are AO corrected over a large enough size to perform accurate photometric calibration.  Single-star/laser adaptive optics systems typically have effective useful fields of arcseconds, which is too small to provide sufficient sources to reliably calibrate a given observation of a SN Ia.  The previously mentioned GeMS instrument achieves correction over 1 arcminute.  A major current challenge in wide-field adaptive optics is to obtain reliable photometric calibration to satisfy the 1\% requirements needed for measuring  accurate distances with Type Ia supernovae. This will require an extension to fields of view with linear sizes of $\approx$10 arcminutes. In addition, further efforts in photometric calibration are necessary to enable Type Ia supernova cosmology to take advantage of these capabilities. Improvements in adaptive optics systems are also important for time-delay-lens cosmography using strongly lensed quasars.

Another approach to adaptive optics, Ground Layer Adaptive Optics (GLAO), concentrates on just the ground-layer contribution to atmospheric turbulence.  In this case there is a trade-off between a PSF improvement over an even larger field of view against correction of just a portion of the turbulent profile.  Existing GLAO systems achieve roughly a factor of 2 improvement in PSF over 10' scale field of view.  Pushing this concept to its limit, it is conceivably possible to correct for the portion of the ground layer turbulence below roughly 100 meters but over the entire $3.5^o$ diameter LSST field of view.  Most existing studies of the ground layer do not separate out this lowest altitude portion of the turbulence, so basic studies of the turbulence are needed to estimate the possible improvement.  We guess that more modest improvements in the PSF of order 20\% are possible. This would still enable a noticeable improvement in highly PSF-dependent effects such as galaxy blending.

\subsection{Fiber Positioner R\&D}

The advent of fiber-fed multi-object spectrographs capable of recording thousands of spectra at a time enabled spectroscopic sky surveys such as BOSS and eBOSS.  Improved spectrographs and the addition of robotically controlled fiber positioners allows the next generation of these surveys to significantly extend the science reach and to measure an order of magnitude more objects.
The best positioner technology for highly multiplexed spectroscopic surveys is still unsettled, with the Prime Focus Spectrograph (PFS), DESI, Multi Object Optical and Near-infrared Spectrograph (MOONS), and the 4-metre Multi-Object Spectroscopic Telescope (4MOST) adopting three different technologies, and the Maunakea Spectroscopic Explorer (MSE) as yet undecided. The Cobra technology adopted for PFS is based on piezoceramic Squiggle motors. It allows an 8mm pitch and maintains fiber telecentricity, but is not under consideration for any future instrument. The phi-theta stepper motor design adopted for DESI and MOONS currently has larger minimum pitch ($\sim$10mm), but the positioning is accurate and repeatable, allowing open-loop `tweaking' of the fiber positions during frame exposure, to compensate for differential refraction across wide-fields. The pitch is limited by commercially available stepper motors, and is likely to reduce in future. The European Southern Observatory MOONS project has shown how patrol radius can be increased, at some cost in coverage uniformity. Swiss, Spanish and US groups continue to refine the design for the MSE project. The Australian Astronomical Observatory (AAO) Echidna tilting spine design used for 4MOST offers the low cost and the smallest pitch (6mm or even less in laboratory studies); it also has the largest patrol radii, giving higher fiber allocation efficiency, and allowing multiple actuator types each with complete sky coverage. The main disadvantage is that the spine tilt introduces defocus and focal ratio degradation, giving 5-10\% losses. The technology continues to be developed for the MSE project; a new low voltage design currently under development is expected to halve the tilt losses, with lower costs and much better modularity. Looking forward to to 2020s, we can expect greatly increased multi-object spectrograph target densities, and/or multiple fibers per target giving kinematic information for each target. 

For lower target densities at the focal plane,  the European Extremely Large Telescope (E-ELT) is proposing autonomous robotic positioners for EAGLE, which allow Multi-Object Adaptive Optics. The AAO has developed the `Starbug' technology, autonomous micro-robots allowing a wide variety of payloads such as deployable Integral Field Units (IFUs). The technology has been adopted for the MANIFEST instrument on the Giant Magellan Telescope (GMT), and it will also be used for the Hector instrument on the 
Australian Astronomical Telescope. For the 2020s, deployable IFUs and multi-object adaptive optics offer the same sort of quantum leap in capabilities that multi-fiber spectroscopy gave in the 1990s.

Fiber positioner systems are just one, albeit a very important, component of multi-object spectrographs. R\&D on other components should also be supported with the goal to maximize the overall system throughout and therefore survey efficiency.

\subsection{Calibration Systems}

For all techniques that use flux measurements as probes of the nature of dark energy, (this includes type Ia supernovae, weak lensing and large scale structure mapping using photometric redshifts, and cluster abundances), improving calibration precision is essential in order to fully exploit the potential of upcoming surveys. This is most apparent in the case of supernova measurements, where we need to precisely compare the photon arrival rates across different passbands in the presence of wavelength-dependent sensor quantum efficiency, optical transmission, and varying atmospheric transmission. The use of type Ia supernovae for cosmology is presently limited by systematic uncertainties in relative flux calibration. But this limitation is not fundamental; both DES and LSST are implementing methods to improve calibration precision. Assuring that colors are measured precisely is also essential for photometric redshift determination. Another important factor that merits further work is the correction for Galactic extinction. Astronomical magnitudes are defined at the top of the Earth's atmosphere, but photons of extragalactic origin suffer scattering as they travel through the Milky Way galaxy. Properly interpreting the apparent intensity of extragalactic sources requires accounting for this attenuation. Finally, in the past few years we have become increasingly aware of subtle features in silicon CCDs that perturb the position, flux, and shape measurements of astronomical sources, as described above. These sensor anomalies can introduce systematic errors in spectroscopic as well as photometric measurements. Understanding and correcting for these sensor anomalies will be vital in next-generation dark energy projects.

\begin{table}
\small
\begin{center}
\begin{tabular}{p{1.5in} p{1.5in} p{3.0in}  }
R\&D Effort &	Goals and Challenges &	Science Topic \\ \hline\hline
CCD Improvements &	Readout noise and speed; Readout architecture; Quantum efficiency and Wavelength coverage & Improved performance for future photometric and spectroscopic surveys. \\ \hline
CMOS Detectors & Pixel-array size; Quantum efficiency; Packaging; Full well; On-chip processing & Applications requiring fast readout such as guiders or multi-object adaptive optics. \\  \hline
Germanium CCDs & Readout noise; pixel-array size & NIR imaging and spectroscopy \\ \hline
Multicolor CCDS & Readout noise; pixel-array size; yield & Low resolution spectroscopy \\ \hline
MKIDs & Pixel-array size, Electronics and data acquisition; Energy resolution; System engineering & Photon counting; Low resolution spectroscopy \\ \hline
21-cm & scale; RFI and foreground systematics control  & Future intensity mapping experiments \\ \hline
Ring Resonator & Device fabrication; Multi-ring devices; On-Sky testing & NIR sky suppression; SN Follow-Up \\ \hline
Adaptive Optics & Performance over large FoV & Reducing PSF of future survey to increase mapping speed and reduce blending\\ \hline
Fiber Positioner & Fiber density; Fiber bundles & Increased throughput for spectroscopic surveys; kinematic lensing; local velocity field; cluster velocity dispersion \\ \hline
Calibration Systems	& Wavelength dependent quantum efficiency, optical transmission and atmospheric effects; Sensor anomalies &	Enhanced performance and improved systematics for future photometric and spectroscopic surveys. \\ \hline
\end{tabular}
\caption{Summary of  R\&D Efforts.}
\label{tab:tech}
\end{center}
\normalsize
\end{table}

\section{Summary and Outlook}

With much of the focus of the DOE's Dark Energy program on DES operations and DESI and LSST construction, it is important to maintain a strong instrumentation and detector R\&D effort to enable the next generation of experiments and to develop the technologies needed for new dark energy probes. Table~\ref{tab:tech} summarizes how the instrumentation program described in this report addresses the science questions raised in the Science Document. Column-wise readout, lower readout noise, single photon counting are just a few examples of improvements that are still possible even for a mature technology like CCDs. New materials such as germanium can enable applications in the near infrared wavelength range. More R\&D is needed for the MKIDs technology to develop its full potential. While 21 cm intensity mapping experiments are currently not part of the DOE's cosmic frontier portfolio, a move in this direction should be considered given the strong science case and the expertise at DOE labs in electronics, computing and instrumentation.
We end this report by pointing out another important aspect of a strong detector development and instrumentation program: it provides essential training for the next generation of experimentalists.
 
\section{Acknowledgments}

Advice and contributions to this report were provided by C. Bebek, J. Estrada, S. Kuhlman, C. Leitz, B. Mazin,  P. O'Connor, A. Parsons, W. Saunders, C. Stubbs, M. Wood-Vasey.





\end{document}